\documentstyle[aps,manuscript]{revtex}
\begin{document}
\draft
\title{Ballistic transport through coupled T-shaped quantum wires}
\author{Yuh-Kae Lin, Kao-Chin Lin and Der-San Chuu\thanks{%
Corresponding author email address: dschuu@cc.nctu.edu.tw;
Fax:886-3-5725230; Tel:886-3-5712121-56105.}}
\address{Department of Electrophysics, National Chiao Tung\\
University \\
1001 Ta Hsueh Road, Hsinchu, 30050 Taiwan}
\date{today}
\maketitle

\begin{abstract}
The ballistic conductance of a coupled $T$-shaped semiconductor quantum wire
(CTQW) are studied. Two types of CTQW are considered, one of which is a $\Pi 
$-shaped quantum wire ($\Pi $QW) which consists of two transverse wires on
the same side of the main wire and the other a $\Pi $-clone quantum wire ($%
\Pi $CQW) which consists of two transverse wires on the opposite sides of
the main wire. The mode matching method and Landauer-Buttiker theory are
employed to study the energy dependence of the ballistic conductance. Most
of transmission profiles of $\Pi $QW and $\Pi $CQW are found to be
distinguishable for large separation $d$ between the two transverse arms.
The transmission probability manifests oscillatory behavior when $d$ is
increased. When a potential is added to the connection region, it results in
decoupling or coupling effects between the two T-shaped wires according to
whether it is positive or negative. When magnetic fields are applied to
CTQW, the transmission profiles are found to be affected profoundly even if
the electrons pass through the field free region only.
\end{abstract}


%\begin{center}

%\end{center}

\preprint{Preprint submitted to Physica B } \vspace{3mm} \narrowtext

\section{Introduction}

Recently, the microetching and epitaxial growth techniques have enabled
semiconductor nanostructures to be fabricated with feature sizes down to
nanometers. Such nanostructures include T-shaped quantum wires in which
quasi-one-dimensional confinement is achieved at the intersection of two
quantum wells. Both experimental and theoretical studies on the nonlocal
ballistic transport of these structures have been stimulated. In general,
T-shaped quantum wires can be fabricated by first growing a GaAs/Al$_{x}$Ga$%
_{1-x}$As multilayers on a (001) substrate, after cleavage, a GaAs quantum
well is grown over the exposed (110) surface, resulting in an array of
T-shaped regions where carrier wavefunctions can be confined in several tens
of angstroms. T-shaped quantum wires (TQW) possess some improved optical
properties of one dimensional excitons, such as the excitonic laser
emission, the enhancement of excitonic binding energy, and the concentrated
oscillator strength. The conductance of such a mesoscopic structure exhibits
many peculiar and interesting features due to its intrinsic nonlocality.
Quantum conductance in mesoscopic structures is the consequence of a complex
scattering process which involves the boundary and the shape of the
potential across the structural geometry as a whole.

Several studies on the electronic transmission properties for a T-shaped
quantum structure have been carried out.$^{1-8}$ \ Many interesting
transmission characteristics, such as resonant transmission and resonant
reflection in the T- shaped structures have been revealed. Such behaviors
are caused from the quantum interference which dominates the ballistic
transport regime. Theoretically, one may view the resonance as being
mediated by the quasibound states of the system. The system of T-shaped
quantum wires has open geometry, therefore, the injected carriers that
travel ballistically over the wire region will across the wire region and
show a strong energy dependent transmission as a consequence of quantum
interference effect induced by the interplay between the propagating modes
of the wires.

By using the scattering matrix approach and Landauer-Buttiker theory,
Goldoni {\it et. al.}$^{4}$ have calculated the conductance of T-shape and
coupled T-shaped quantum wires with different wire widths. The transmission
coefficient of the whole coupled T-shaped quantum wires can be obtained
easily since the total T-matrix is the product of the T-matrices of the
isolated wires. The double resonance obtained in their result is ascribed to
a fingerprint of the bonding and antibonding combinations of the resonance
states of the isolated wires. Bohn$^{5}$ has introduced a periodic array of
T-shaped devices. He showed that deflected arrays exhibit a unique resonance
structure with respect to electrons traveling along the array. The
coefficients of the reflection and transmission through the array can peak
simultaneously at resonance. Unlike the analogous case in superlattices, the
peaks are at energies where the wavelength $\lambda $ satisfies the
condition $n\lambda /2=d$ for some integer $n$. Consequently, the scattering
wave function possesses nodes at the intersection of the longitudinal arm
and the transversal arm, and thus greatly reduces the flux lost to
transversal leads. Nikolic and Sordan$^{6-7}$ have also studied the
transmission properties of a quantum waveguide system with attached stubs in
the ballistic regime. They found the transconductance and the differential
drain conductance are small. Their result suggests limited abilities for
conventional application of the transistor. Chen {\it et. al.}$^{7}$.
calculated transmission of electrons in a T-shaped opened quantum waveguide
(TOQW) subjected to an inhomogeneous magnetic field perpendicular to the
TOQW plane with mode-matching technique. The transmission profiles are found
to depend sensitively on geometric parameters.

In this work, we study a $\Pi $-shaped opened quantum structure and its
clone shape, which are four-terminal waveguide-like structures,
schematically as shown in Fig.1. We take first the geometric variation into
account. Second, the interconnection region is considered to be acted by a
potential. Third, the magnetic field is considered to apply to the vertical
wires. Unlike the stubs, arms of the structures considered in our case are
assumed to be long enough and open in the longitudinal and the transverse
directions. The centers of the two vertical arms are spaced by a distance $d$
as shown in Fig.1. The scattering matrice are calculated by using
mode-mathing method. Our model will be presented briefly in the next
section. Results and discussions will be given in the final section.\newline

\section{Model and Formalism}

\ \ \ \ \ We model the structure geometry as illustrated in Fig.1: A
horizontal wire with a width of $W_{1}$ $(W_{2})$ for region I (VII), a
vertical wire with a width of $W_{2}$ $(W_{3})$ for region II (V), an
interconnection part for region IV and a junction region with an area of $%
W_{1}\times W_{2}$ $(W_{1}\times W_{3})$ for region III (VI). In the wire,
2DEG system with perfect barrier confinement (e.g. high quality interfaces)
is assumed. The individual electron propagates ballistically through the
entire wire. The transverse potential inside the wire is set to zero. The
Schr\"{o}dinger equation of individual electron can be written as 
\begin{equation}
-\frac{\hbar ^{2}}{2m}\nabla ^{2}\Psi =E\Psi   \label{1}
\end{equation}%
\ \ \ The whole quantum wire can be split into several individual
homogeneous subregions: horizontal region I, vertical regions II and V,
intersection region III and VI, interconnecting region IV, and the outgoing
horizontal region VII. The two intersection regions act as scattering
centers. And the interconnecting region acts as a connection of the two
TQWs. An $n$th mode electron is considered to inject from left of region I
into the wire. The wave function in region I can be written in terms of a
sum of incident and reflecting modes as 
\begin{equation}
\Psi _{n}^{I}(x,y)=\Phi _{n}^{I\left( +\right) }\left( y\right)
e^{ik_{n}^{I\left( +\right) }\left( x+0.5W_{2}\right) }+\sum_{m}R_{mn}\Phi
_{m}^{I\left( -\right) }\left( y\right) e^{ik_{m}^{I\left( -\right) }\left(
x+0.5W_{2}\right) },  \label{2}
\end{equation}

where $k_{n}^{I\left( \pm \right) }=\sqrt{k^{2}-\left( n\pi /W_{1}\right)
^{2}},\ \pm $ represents the incident or reflecting mode, respectively, and $%
\Phi _{n}^{I(\pm )}$ are envelope functions in region I. The wave functions
in regions II , V, and VII are given by a sum of outgoing modes
respectively, i.e.,

\begin{equation}
\Psi _{n}^{II}\left( x,y\right) =\sum_{m}S_{mn}^{\left( 1\right) }\Phi
_{n}^{II\left( +\right) }\left( x\right) e^{ik_{m}^{II\left( +\right)
}\left( y-0.5W_{1}\right) },  \label{3}
\end{equation}

\begin{equation}
\Psi _{n}^{V}\left( x,y\right) =\sum_{m}S_{mn}^{\left( 2\right) }\Phi
_{m}^{V\left( \pm \right) }\left( x\right) e^{ik_{m}^{V\left( \pm \right)
}\left( y\mp 0.5W_{1}\right) },  \label{4}
\end{equation}

where $\pm $ represents the upward or downward arm and

\begin{equation}
\Psi _{n}^{VII}\left( x^{\prime },y\right) =\sum_{m}T_{mn}\Phi
_{m}^{VII\left( +\right) }\left( y\right) e^{ik_{m}^{VII\left( +\right)
}\left( x^{\prime }-0.5W_{3}\right) }.  \label{5}
\end{equation}

The wave function in region IV is given by the sum of rightgoing $\left(
+\right) $ and leftgoing $\left( -\right) $ modes,

\begin{equation}
\Psi _{n}^{IV}\left( x,y\right) =\sum_{m}\left[ U_{mn}\Phi _{m}^{IV\left(
+\right) }\left( y\right) e^{ik_{m}^{IV\left( +\right) }\left(
x-0.5W_{2}\right) }+V_{mn}\Phi _{m}^{IV\left( -\right) }\left( y\right)
e^{ik_{m}^{IV\left( -\right) }\left( x-0.5W_{2}\right) }\right] .  \label{6}
\end{equation}

In region III and region VI, all modes must be taken into account, thus

\begin{eqnarray}
\Psi _{n}^{III}\left( x,y\right) &=&\sum_{j}f_{j}\left( y\right) \cdot \left[
a_{jn}\sin \left( k_{j}^{\prime }\left( x-0.5W_{2}\right) \right)
+b_{jn}\sin \left( k_{j}^{\prime }\left( x+0.5W_{2}\right) \right) \right] 
\nonumber \\
&&+\sum_{j}g_{j}\left( x\right) c_{jn}\sin \left( k_{j}^{\prime \prime
}\left( y+0.5W_{1}\right) \right) ,  \label{7}
\end{eqnarray}

\begin{eqnarray}
\Psi _{n}^{VI}\left( x^{\prime },y\right) &=&\sum_{j}f_{j}\left( y\right)
\cdot \left[ d_{jn}\sin \left( k_{j}^{\prime }\left( x^{\prime
}-0.5W_{3}\right) \right) +e_{jn}\sin \left( k_{j}^{\prime }\left( x^{\prime
}+0.5W_{3}\right) \right) \right]  \nonumber \\
&&+\sum_{j}g_{j}^{\prime }\left( x^{\prime }\right) h_{jn}\sin \left(
k_{j}^{\prime \prime \prime }\left( y\pm 0.5W_{1}\right) \right) .  \label{8}
\end{eqnarray}
Here $f_{j}\left( y\right) =\sqrt{\frac{2}{W_{1}}}\sin \left( \frac{j\pi }{%
W_{1}}\left( y+0.5W_{1}\right) \right) ,$ $g_{j}\left( x\right) =\sqrt{\frac{%
2}{W_{2}}}\sin \left( \frac{j\pi }{W2}\left( x+0.5W_{2}\right) \right) $ and 
$g_{j}^{\prime }\left( x^{\prime }\right) =\sqrt{\frac{2}{W_{3}}}\sin \left( 
\frac{j\pi }{W3}\left( x^{\prime }+0.5W_{3}\right) \right) $ represent the
transverse wave functions of the electron in mode $j$ \ inside the different
regions of the wires, and are used as the expansion bases. The wave numbers $%
k_{j}^{\prime }=\sqrt{k^{2}-\left( j\pi /W_{1}\right) ^{2}},$ $k_{j}^{\prime
\prime }=\sqrt{k^{2}-\left( j\pi /W_{2}\right) ^{2}},$ and $k_{j}^{\prime
\prime \prime }=\sqrt{k^{2}-\left( j\pi /W_{3}\right) ^{2}}$ are either real
for propagating modes or pure imaginary for evanescent modes. Now expand the
wavefunctions in terms of a set of complete bases corresponding to the
transverse eigenfunctions in regions I, II, IV, V and VII, respectively as 
\begin{equation}
\Phi _{n}^{I(\pm )}(y)=\sum_{j}\alpha _{jn}^{I(\pm )}f_{j}(y),  \label{9}
\end{equation}

\begin{equation}
\Phi _{n}^{II(+)}(x)=\sum_{j}\beta _{jn}^{II(+)}g_{j}(x),  \label{10}
\end{equation}
\begin{equation}
\Phi _{m}^{IV(\pm )}(y)=\sum_{j}\gamma _{jm}^{IV(\pm )}f_{j}(y),  \label{11}
\end{equation}
\begin{equation}
\Phi _{n}^{V(\pm )}(x^{\prime })=\sum_{j}\delta _{jn}^{V(\pm )}g_{j}^{\prime
}(x^{\prime }),  \label{12}
\end{equation}
and 
\begin{equation}
\Phi _{n}^{VII(+)}(y)=\zeta _{jn}^{VII(+)}f_{j}(y).  \label{13}
\end{equation}
Substituting these functions into Eq.(1) for a given Fermi energy $E_{F}$,
we obtain five sets of eigen-wave-numbers $\{k_{n}^{I(\pm )}\}$, $%
\{k_{n}^{II(+)}\}$, $\{k_{n}^{IV(\pm )}\}$ $\{k_{n}^{V(\pm )}\}$, and $%
\{k_{n}^{VII(+)}\}$ and eigen-wave-functions $\{\Phi _{n}^{I(\pm )}(y)\}$, $%
\{\Phi _{n}^{II(+)}(y)\}$, $\{\Phi _{n}^{IV(\pm )}(y)\}$, $\{\Phi
_{n}^{V(\pm )}(y)\}$, and $\{\Phi _{n}^{VII(+)}(x)\}$. By using boundary
matching technique,$^{9}$ we can derive all coefficients in Eqs. (2)--(8)
such as $\{r_{mn}\}$, $\{s_{mn}^{\left( 1\right) }\}$, $\{s_{mn}^{\left(
2\right) }\}$, $\{u_{mn}\}$, $\{v_{mn}\}$, $\{t_{mn}\}$,$\{a_{jn}\}$, $%
\{b_{jn}\}$, $\{c_{jn}\}$.$\{d_{jn}\}$, $\{e_{jn}\}$, and $\{h_{jn}\}$.

The group velocities of the $j$ th state in region I, II,\ V and VII are
respectively 
\begin{equation}
V_{j}^{I(\pm )}=\frac{\hbar }{m^{\ast }}\int_{-0.5W_{1}}^{0.5W_{1}}\Phi
_{j}^{I(\pm )}(y)k_{j}^{I(\pm )}\Phi _{j}^{I(\pm )}(y)dy,  \label{14}
\end{equation}
\begin{equation}
V_{j}^{II(+)}=\frac{\hbar }{m^{\ast }}\int_{-0.5W_{2}}^{0.5W_{2}}\Phi
_{j}^{II(+)}(x)k_{j}^{II(+)}\Phi _{j}^{II(+)}(x)dx,  \label{15}
\end{equation}
\begin{equation}
V_{j}^{V(\pm )}=\frac{\hbar }{m^{\ast }}\int_{-0.5W_{3}}^{0.5W_{3}}\Phi
_{j}^{V(\pm )}(x)k_{j}^{V(\pm )}\Phi _{j}^{V(\pm )}(x)dx,  \label{16}
\end{equation}
as well as 
\begin{equation}
V_{j}^{VII(+)}=\frac{\hbar }{m^{\ast }}\int_{-0.5W_{1}}^{0.5W_{1}}\Phi
_{j}^{VII(+)}(y)k_{j}^{VII(+)}\Phi _{j}^{VII(+)}(y)dy.  \label{17}
\end{equation}

The transmission probabilities $\widetilde{t}_{nj}$ (in region VII ) and $%
\widetilde{s}_{nj}^{\left( 1\right) }\left( \widetilde{s}_{nj}^{\left(
2\right) }\right) $ (in region II and region V) from the incident mode $n$
to the final mode $j$ \ , and the reflection probability $\widetilde{r}_{nj}$
from the incident mode $n$ to the final mode $j$ (in region I ) can be
obtained, respectively, as follows: 
\begin{equation}
\widetilde{r}_{nj}=\frac{V_{j}^{I(-)}}{V_{n}^{I(+)}}\mid r_{nj}\mid ^{2},
\label{18}
\end{equation}
\begin{equation}
\widetilde{s}_{nj}^{\left( 1\right) }=\frac{V_{j}^{III(+)}}{V_{n}^{I(+)}}%
\mid s_{nj}^{\left( 1\right) }\mid ^{2},  \label{19}
\end{equation}
\begin{equation}
\widetilde{s}_{nj}^{\left( 2\right) }=\frac{V_{j}^{III(+)}}{V_{n}^{I(+)}}%
\mid s_{nj}^{\left( 2\right) }\mid ^{2},  \label{20}
\end{equation}
\begin{equation}
\widetilde{t}_{nj}=\frac{V_{j}^{II(+)}}{V_{n}^{I(+)}}\mid t_{nj}\mid ^{2}.
\label{21}
\end{equation}
It should be emphasized that the expansions (9)--(13) involve an infinite
sum including all possible evanescent modes. In practice, in order to solve
this set of equations numerically, we have to truncate the sum at some
finite number which should be large enough to achieve a desired accuracy.
The numerical convergence can be checked by flux conservation. The
relationship $\sum_{j}(\widetilde{t}_{jn}+\widetilde{r}_{jn}+\widetilde{s}%
_{jn})=1$ should be fulfilled accurately.

The total transmission coefficients T and S are then given by 
\begin{equation}
T=\sum_{n=1}^{N_{1}}\sum_{j=1}^{N_{2}}\widetilde{t}_{nj},  \label{22}
\end{equation}
\begin{equation}
S=\sum_{n=1}^{N_{1}}\sum_{j=1}^{N_{3}}\widetilde{s}_{nj}.  \label{23}
\end{equation}
Where $N_{1}$, $N_{2}$ and $N_{3}$ are the numbers of propagating modes in
regions I, II and III, respectively. The conductance G at zero temperature
is given by the Landauer--Buttiker formula: 
\begin{equation}
G_{t}=(2e^{2}/h)T  \label{24}
\end{equation}
and 
\begin{equation}
G_{s}=(2e^{2}/h)S.  \label{25}
\end{equation}
We also evaluate the probability density of electrons in the quantum wire by
adding the contributions from all propagating modes as 
\begin{equation}
\rho (x,y)=\sum_{n=1}^{N}\mid \Psi _{n}(x,y)\mid ^{2}/k_{n}.  \label{26}
\end{equation}

\section{Numerical Results and Discussions}

\subsection{Transmission Properties with Geometric Variations}

Figs.1(a) and (b) schematically depict the geometry of the $\Pi $--shaped QW
($\Pi $QW) and the $\Pi $-clone QW ($\Pi $CQW). We present our results in
terms of some convenient parameters: 1) the first threshold energy $E_{1}=%
\frac{\hbar ^{2}}{2m^{*}}(\pi /W_{1})^{2}$ through horizontal wire, 2) the
distance $d$ between two center of the intersections of vertical wires and
horizontal wire, 3) the ratios of widths $\alpha =W_{2}/W_{1}$ and $\gamma
=W_{3}/W_{2}$.

First of all, we consider that all wires are the same width, namely $W$.
Transmission probabilities are calculated with varying $k_{F}$ as shown in
Fig. 2(a) and (b) for different $d$. Curves from bottom to top in Fig.2(a)
are shifted by 1.0 for clarity and correspond to the cases of $d=$ 1, 1.1,
1.2, 1.3, 1.5, 1.7, and 2.0$W_{1}$, respectively. And curves from bottom to
top in Fig.2(b) are shifted by 1.0 for clarity and correspond to the cases
of $d=$ 2, 2.5, 3.0, and 5.0$W_{1}$, respectively. Hereafter, we present the
transmission probabilities of the $\Pi $QW system as solid lines and those
of the $\Pi $CQW system as the dotted lines in all figures.

For $d=$ 1 , the vertical wires are adjacent to each other. Thus, a $\Pi $QW
with $d=1$, can be regarded as a TQW with a double width in the vertical arm
except there is an infinite thin wall along the vertical arm axis. However,
one can note from the figure that the profiles of transmission of a $\Pi $QW
with $d=1$ are quite different from the transmission profiles of a TQW with
the same width ($2.0W_{1})$\ of the vertical arm$^{8}$. \ In fact, the
bottom curve of $\Pi $QW ( $d=1)$ is similar to the result obtained in TQW
with a vertical arm of $1.0W_{1}$ width as obtained in Ref (8). This implies
that the two systems are similar except the transmission amplitude is
suppressed in a $\Pi $QW system. For the $\Pi $CQW, the sharp dip at $%
k_{F}=2.0\,\pi /W_{1}$ is replaced by a wider valley before $k_{F}=2.0\,\pi
/W_{1}$. The transmission behaviors of the two structures ( $\Pi $QW and $%
\Pi $CQW ) are different in general, however, their periodic oscillations
are the same. The period of the oscillation is dominated by the distance $d$
as can be seen from Fig.2. The periodicity can be fitted as $n\lambda
_{l}=2d $ approximately, where $n$ is the number of periods in one mode, and 
$\lambda _{l}=2\pi /(k_{F}-\pi /W_{1})$ denotes the longitudinal wave length
of the incident electron waves. Thus, once one finds two peaks in the region 
$1.0<k_{F}W_{1}/\pi <2.0$ for $d=\,1W_{1}$, then four peaks will be found
for $d=\,2W_{1}$, and so on. Curves in both structures ( $\Pi $QW and $\Pi $%
CQW ) possess peak-dip structures. Especially, these peak-dip structures are
more clear for larger $d$ at $k_{F}=2.0\,\pi /W_{1}$. On the contrary, they
are observed only in certain circumstance for smaller $d$. According to the
previous result$^{8}$, there exists a localized state in the intersection
region for a symmetric TOQW with same wavenumber $k_{F}=2.0\,\pi /W_{1}$.
The peak-dip structure at $k_{F}=2.0\pi /W_{1}$ can be ascribed to this
localized state. The peak-dip structure is found at $k_{F}=2.0\,\pi /W_{1}$
on the curve with $d=1.5$ for $\Pi $CQW. For $d$ larger than $1.5\,W_{1}$,
the peak-dip structure is sharper in $\Pi $CQW than that in $\Pi $QW.

Due to the fact that both $\Pi $QW and $\Pi $CQW structures are equivalent
to a system of two TQWs, one may expect that the transmission properties of
these two structures will be the same if the coupling between the
constituent TQWs becomes very weak. However, our result does not manifest
this accordance. On the contrary, the two transmission profiles are still
distinguishable from each other even for large $d$. It is also found that
the transmission probabilities vary periodically with $d$ for a fixed
wavenumber as shown in Fig.3. These behaviors are the essential
characteristics of ballistic theory.

Now let us consider the case that the widths of the vertical wires are the
same, while the ratio of the width of the vertical wire to the horizontal
wire is varied. The result is displayed in Fig. 4. For simplicity, we define
the ratio of the width of the vertical wire to the horizontal wire as $%
\alpha =W_{2}/W_{1}=W_{3}/W_{1}$. And the distance $d$ is set to $2\,\alpha
W_{1}$. Curves from the bottom to the top are shifted by 1.0 individually
for clarity and correspond to the cases of $\alpha $ =0.1, 0.2, 0.3, 0.5,
0.7, 1.0, 1.5, 2.0 and 4.0, respectively, as shown in the figure. For
extremely small $\alpha $, perfect stepwise profiles are observed in both
structures. The transmissions are strongly suppressed when the ratio $\alpha 
$ is large (e.g. 2.0 and 4.0). The solid curve for $\alpha =0.5$ agrees with
the result of previous work$^{4}$. A double resonance is evident either on
the curve of $\alpha =0.3$ or the curve of $\alpha =0.5$. They are the
signature of the bonding and antibonding combinations of the resonant
quasi-1D state of isolated wires.

Finally, the transmission profiles in the $\Pi $QW and $\Pi $CQW with
vertical wires of different width are considered. For simplicity, the width
of one vertical wire is kept to be the same as that of the horizontal wire.
The calculated transmission profiles for $W_{1}=W_{3}$ and different $W_{2}$
are shown in Fig.5(a) and those for $W_{1}=W_{2}$ \ and various $W_{3}$ are
shown in Fig.5(b). Curves are offset for clarity. Curves from bottom to top
correspond to the cases of $\alpha $=0.1, 0.2, 0.3, 0.5, 0.7, 1.0, 2.0, 4.0,
and 5.0, respectively. Where $\alpha $ is the ratio of the wire width of the
vertical wire to the horizontal wire. It is observed that the transmission
probability is drastically suppressed for large $\alpha $ as can be seen
from the upper curves in (a) and (b). Comparing curves in (a) and (b), we
observe that the transmission profiles are the same. When $\alpha $ is
small, the transmission profiles of the $\Pi $QW and $\Pi $CQW become
indistinguishable and almost the same as that of TQW system. The double
resonance can be observed again.

\subsection{Transmission Under an Additional Potential}

We now consider the case that an additional scalar potential is applied to
the interconnection region IV. The applied potential can be negative or
positive for attracting or depleting electrons. The different coupling
profiles are interesting and may be important for practical usage of the
mesoscopic devices.

Fig.6 presents the calculated transmission profiles for different potential
strength $V_{4}$ in unit of $E_{1}$. Here we consider $W_{3}=W_{2}=W_{1}$,
and $d=2W_{1}$. Figs.6(a) and (c) correspond to the positive potential for
electrons. Figs.6(b) and (d) correspond to negative potential. Curves are
offset for clarity. As shown in Fig.6(a), one can observe that the positive
potential does not affect the transmission very much when $V_{4}\leq E_{1}$.
>From Figs.6(a) and (c), two features are shown: (1) the onset is shifted
due to the depletion potential; (2) the positions of transmission dips are
not changed. On the contrary, Figs.6(b) and (d) show that the additional
negative potential affects the transmission much stronger than the positive
one. Especially, the potential enhances the coupling between the two TOQWs
as one can note from the fact that the resonant dip-peak-dip structure
becomes broader and shallower when the potential is increased. More peaks
are on the curves and the positions of dips are not changed as the case of
positive potential. Moreover, it can be observed that discrepancy between
the two structures becomes prominent as the potential strength is increased.
These results manifest that the negative potential increases the coupling
strength between the two individual TOQWs.

\subsection{Transmissions under the influence of surrounding magnetic Fields}

Finally, magnetic fields are considered to apply to the vertical wires only,
therefore, the electrons pass through the main arm regions with no
additional field. We shall study the effect of the surrounding magnetic
fields on the transmission behavior. First, we consider the magnetic field
is applied only to one of the vertical wires, i.e on arm II or arm V. The
direction of the field is perpendicular to the 2DEG plane. Transmission
probabilities are calculated as a function of Fermi wave vector as depicted
in Fig.7. Curves in Fig.7(a) are offset for clarity, and correspond to the
cases of magnetic field strength $B=$ -1.0, -0.7, -0.5, -0.3, -0.2, -0.1,
0.1, 0.2, 0.3, 0.5, 0.7, 1.0 Tesla, respectively in region II. Those shown
in Fig.7(b) are the same except the magnetic field is applied to region V.
>From these curves, one can conclude that: (1) the magnetic field does
affect the transmission, although the electrons do not pass through the
region with magnetic field directly. This phenomenon is according with
Aharonov-Bohm effect. However, no periodic behavior can be found. (2) For
the $\Pi $QW system as shown in solid curves in Figs.7(a) and (b), both
cases show a one-to-one correspondence to each other. This manifests that
the influence of the magnetic field on the transmission profile depends only
on the magnetic field strength. However, there is no correspondence in the
case of $\Pi $CQW which is presented by the dotted lines in Figs.7(a) and
(b). (3) Generally speaking, opposite polarity of the magnetic field causes
different influence on the transmission in $\Pi $CQW systems.

The transmission profiles versus Fermi wave number $k_{F}$ for the case that
the magnetic field being applied to both regions II and V, are displayed in
Fig.8. Fig.8(a) presents the transmission in the $\Pi $QWs and $\Pi $CQWs
with same polarity in both vertical arms, and Fig.8(b) presents those with
opposite polarity to each other in the two vertical arms. The curves are
offset for clarity. The solid lines represent the $\Pi $QW systems and
dotted lines represent the $\Pi $CQWs systems. It is found that for the case
of $\Pi $QWs, though the geometry and the applied field are symmetric, the
transmission probabilities are different from each other (e.g. the solid
curves with $v=0.2$ and $-0.2$) as can be seen from Fig.8(a). However, for
the case of $\Pi $CQWs, the transmission are polarity independent as can be
noted from the dotted curves in Fig.8(a). No such symmetry can be found in $%
\Pi $CQW as shown in Fig.8(b). Furthermore, peak-dip structures are evident
both in Figs.(a) and (b) at high field situations, though the electrons
always move in field free region. One can expect that the transmission
profiles will become stepwise structures when the applied magnetic field is
extremely high. And in the intermediate field strength, the magnetic field
changes the oscillatory behavior of the profiles significantly.

\section{summary}

In the present work, the transmission properties of the coupled TOQWs are
found to be very sensitive to the geometric configurations as well as the
strength and polarity of the applied fields. A double resonance is observed
on the profiles at certain ratio of the width of the vertical wire to the
horizontal wire. The transmission is suppressed drastically as the width of
one or both vertical wires become large. Most of the transmission profiles
of $\Pi $QW and $\Pi $CQW are distinguishable even for large inter- distance 
$d$ between the two vertical wires. The\ transmission profiles exhibit
oscillatory behavior as the distance $d$ is increased and manifest periodic
features as the distance $d$ is varied. T-shaped quantum wires have been
proposed to achieve the quantum interference effect by controlling the
length of its lateral closed arms. In the present study, it is found that
the interference pattern can be easier to obtain by modulating the length
and width of transversal arms and the distance between arms.

When a potential is added to the connection region, it results in decoupling
or coupling effects between the two TQWs according to whether it is positive
or negative. This behavior is observed by the alternating occurrence of the
successive dips and valleys when the potential is increased positively.

Though the electrons pass through only the field free region, the magnetic
field still affects the transmission in the QWs profoundly. The perfect
transmission can be seen only in the high magnetic field region.

Acknowledgement: This work is supported partially by National Science
Council, Taiwan under the grant number NSC91-2112-M-009-002.

\begin{figure}[tbp]
\caption{(a) The schematic illustrations of the geometries of a $\Pi $QW
system. (b) a $\Pi $CQW system.}
\label{fig:pgeo}
\end{figure}
\begin{figure}[tbp]
\caption{The transmission $T\ $ versus $K_{F}$ for different $\ d\ $which is
converted to $v$ by $d=(1+v)W_{1\text{ }}$. \ All wires have same width. (a) 
$d\leq 2W_{1}$ (b)$d\geq 2W_{1}$. The solid lines represent the $T$ of $\Pi $%
QW, while the dotted lines represent the $T$ of $\Pi $CQW.}
\label{fig:pt}
\end{figure}
\begin{figure}[tbp]
\caption{The periodic behaviors of transmissions versus $d$ for $%
W_{3}=W_{2}=W_{1}.$ $T_{1}(T_{2}),S_{1}(S_{2})$ represent the total
transmission coefficients $T$ and $S$ as defined in Eqs.(22) and (23)\ for $%
\Pi $QW ( $\Pi $CQW) .}
\label{fig:pperiodic}
\end{figure}

\begin{figure}[tbp]
\caption{ Transmission versus $k_{F}$ for different $\protect\alpha $. Where 
$\protect\alpha $ is the ratio of the width of the vertical arm to the
horizontal arm. And $d=$ $2\protect\alpha W_{1}$. The solid lines represent
the $T$ of $\Pi $QW, while the dotted lines represent the $T$ of $\Pi $CQW.
Curves are offset for clarity.}
\label{fig:pdi}
\end{figure}

\begin{figure}[tbp]
\caption{Same as Fig.2, except the width of one vertical arm varies. (a) The
width of $W_{2}$ varies, (b) the width of $W_{3}$ varies. The solid lines
represent the $T$ of $\Pi $QW, while the dotted lines represent the $T$ of $%
\Pi $CQW.}
\label{fig:pw}
\end{figure}

\begin{figure}[tbp]
\caption{Transmission profiles versus $k_{F}$ for a potential $V_{4}$
applied to the region IV. (a) and (c) correspond to the positive potential
and (b) and (d) correspond to the negative potential. The solid lines
represent the $T$ of $\Pi $QW,while the dotted lines represent the $T$ of $%
\Pi $CQW.}
\label{fig.v4}
\end{figure}

\begin{figure}[tbp]
\caption{Transmission profiles versus $k_{F}$ for the magnetic field applied
to only one vertical arm. (a) to region II, and (b) to region V. The solid
lines represent the $T$ of $\Pi $QW,while the dotted lines represent the $T$
of $\Pi $CQW. Curves are offset for clarity.}
\label{fig:pb25}
\end{figure}

\begin{figure}[tbp]
\caption{Transmission profiles versus $k_{F}$ for the magnetic field applied
to both vertical arms. (a) same polarity, and (b) opposite polarity in II
and V. The solid lines represent the $T$ of $\Pi $QW, while the dotted lines
represent the $T$ of $\Pi $CQW. Curves are offset for clarity.}
\label{fig:pbb}
\end{figure}

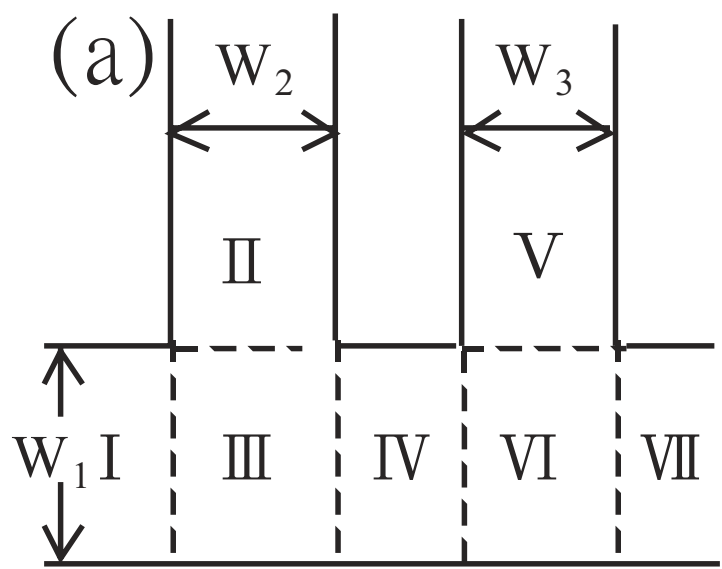
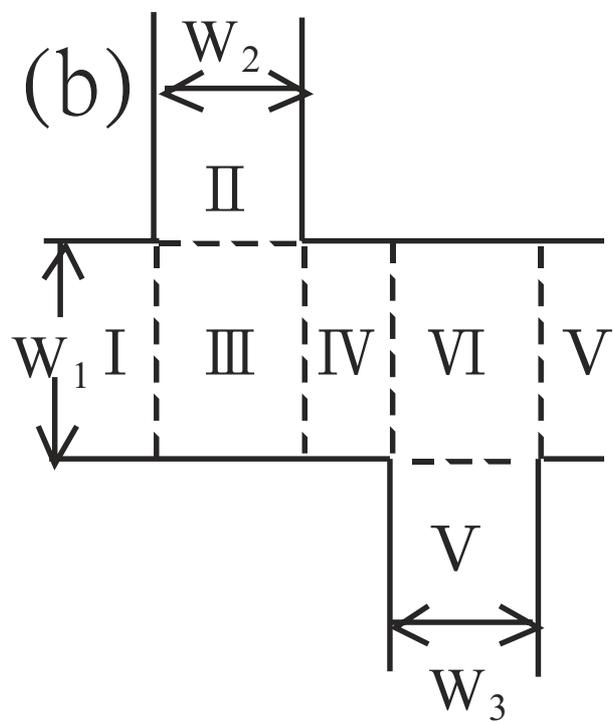
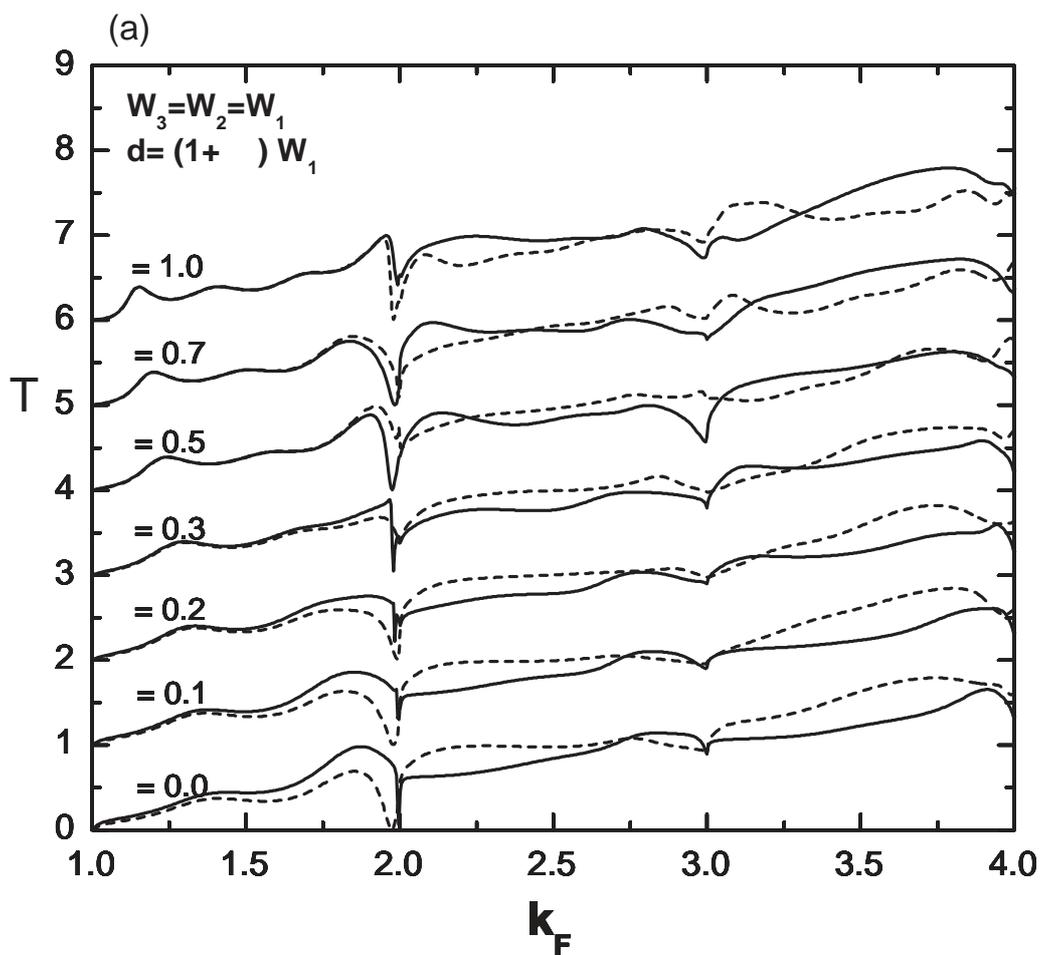
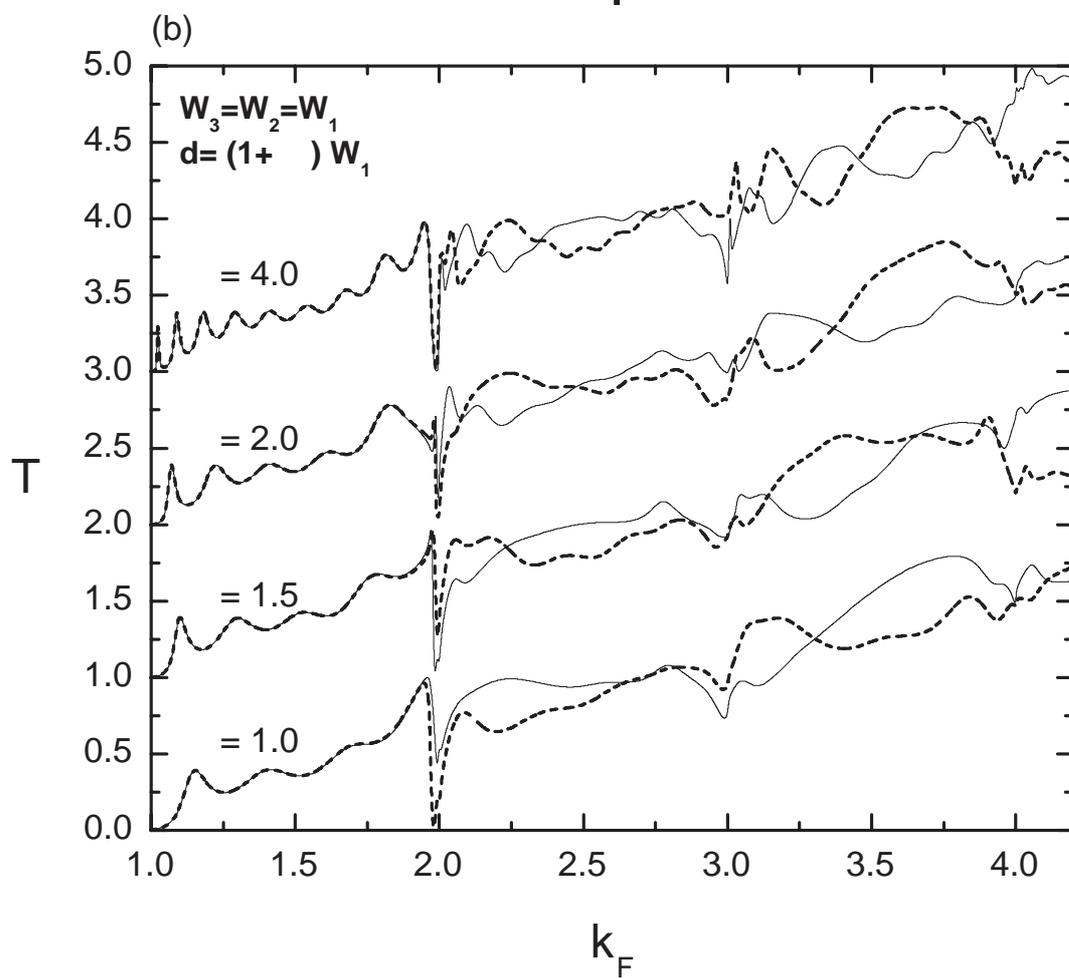
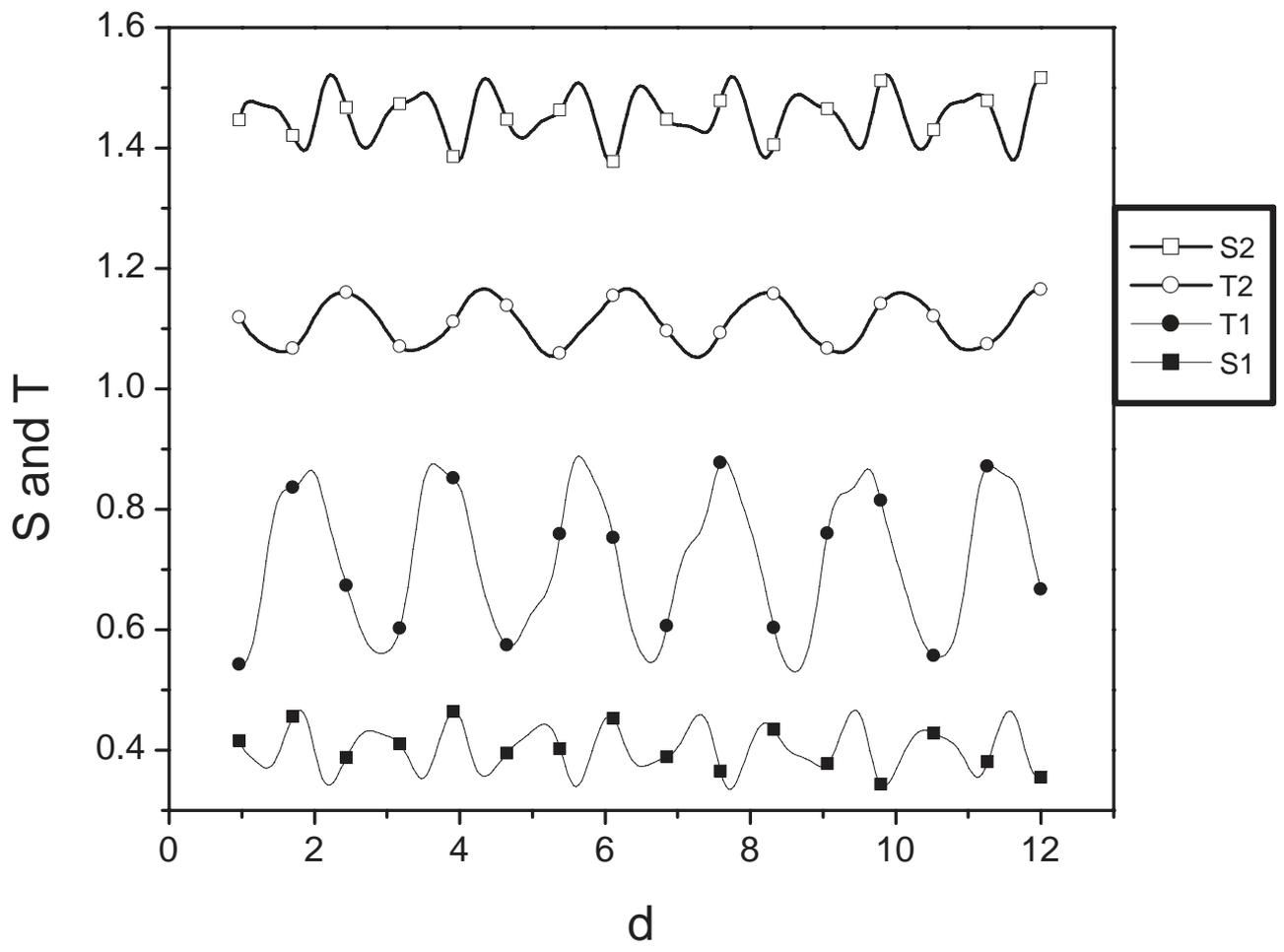
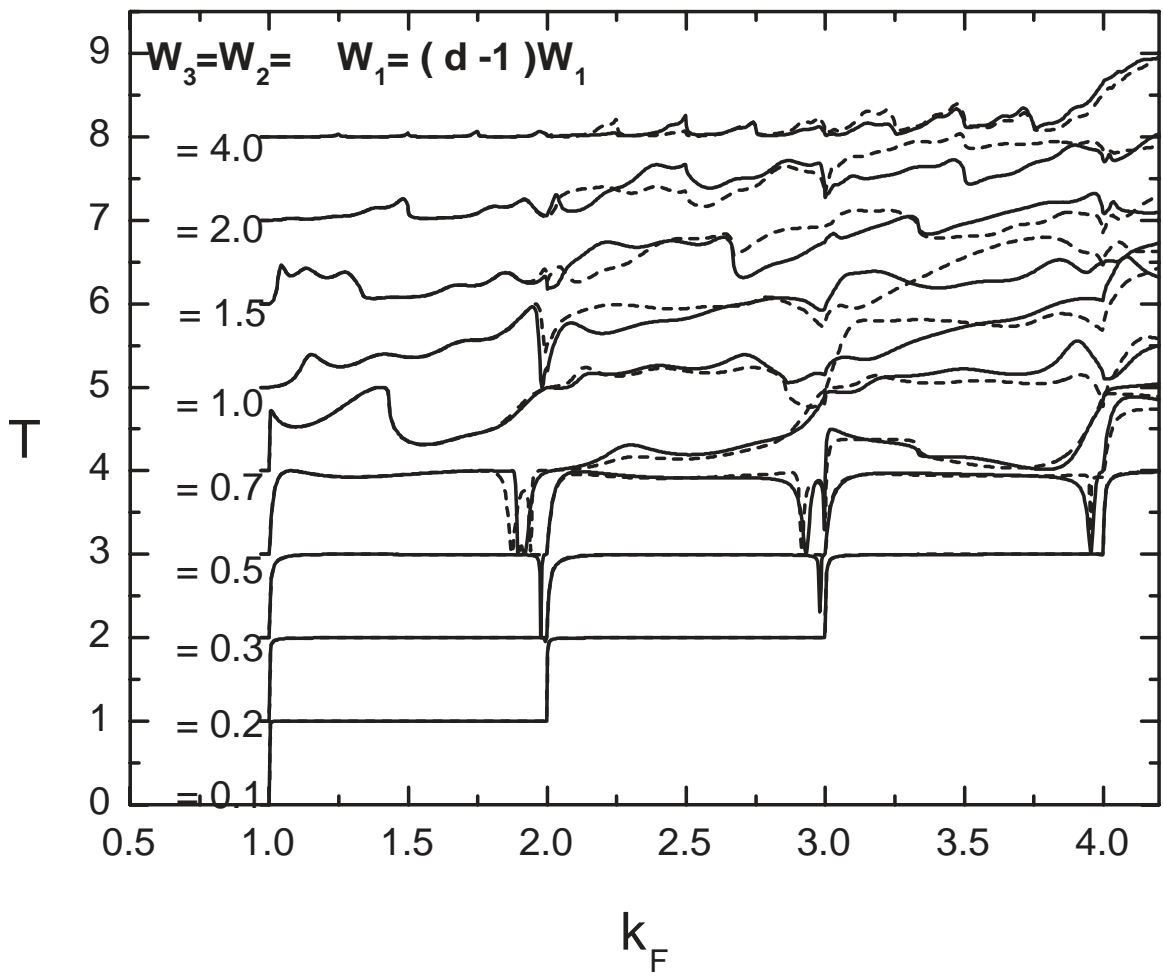
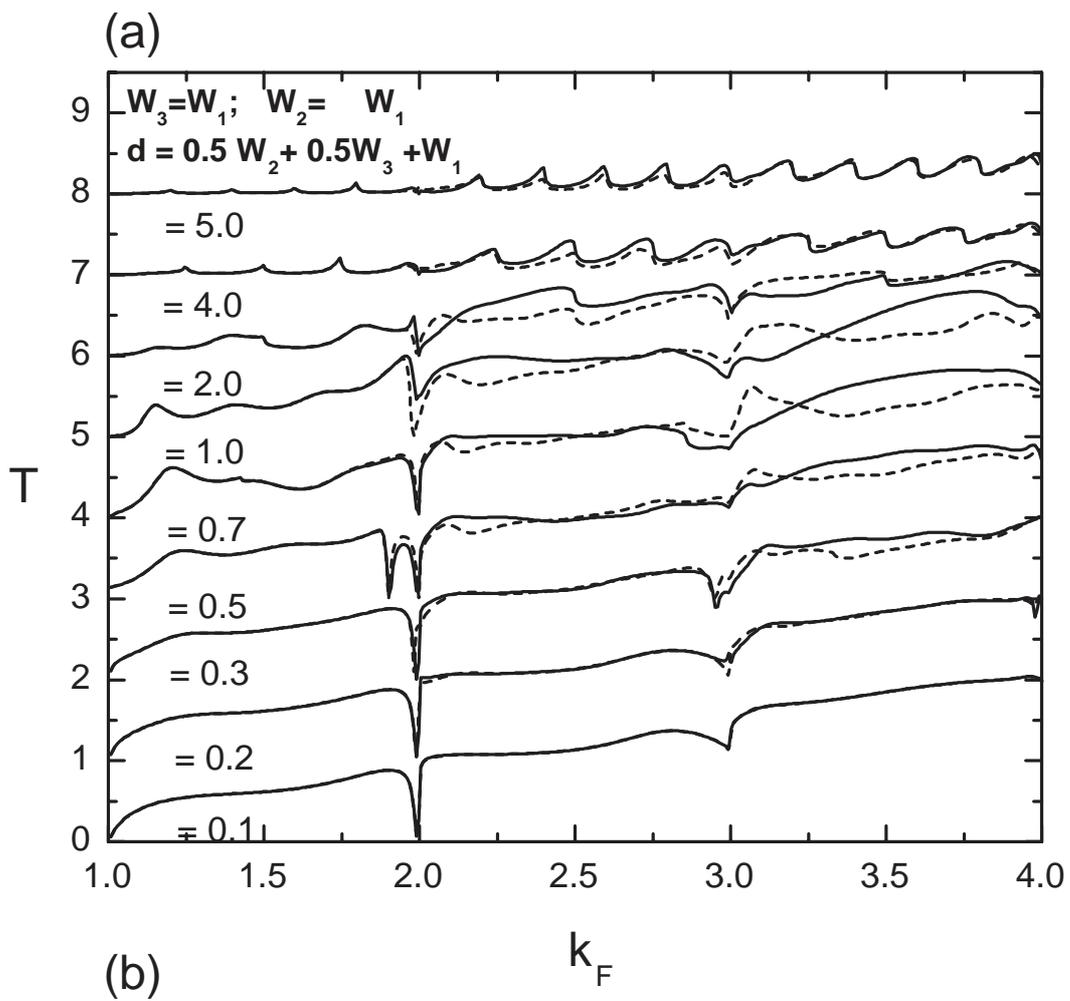
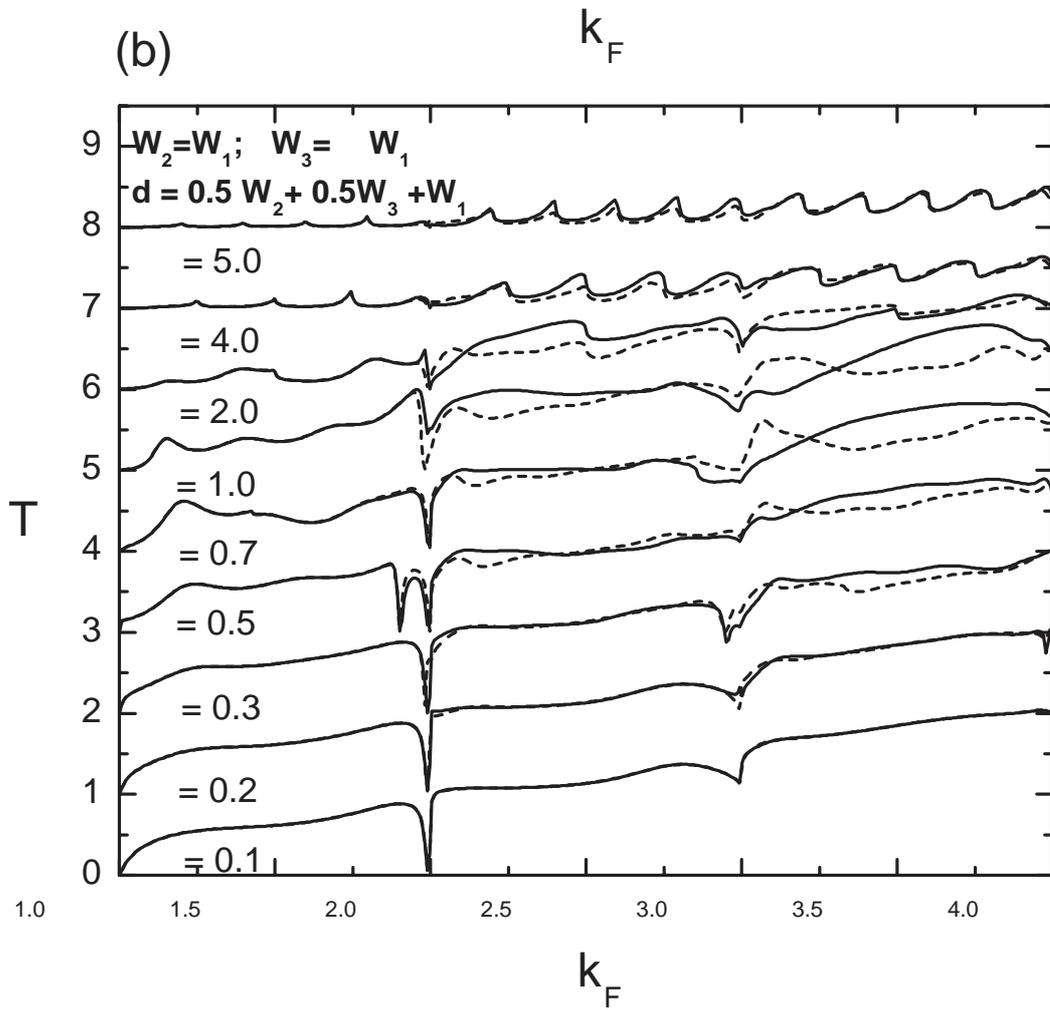
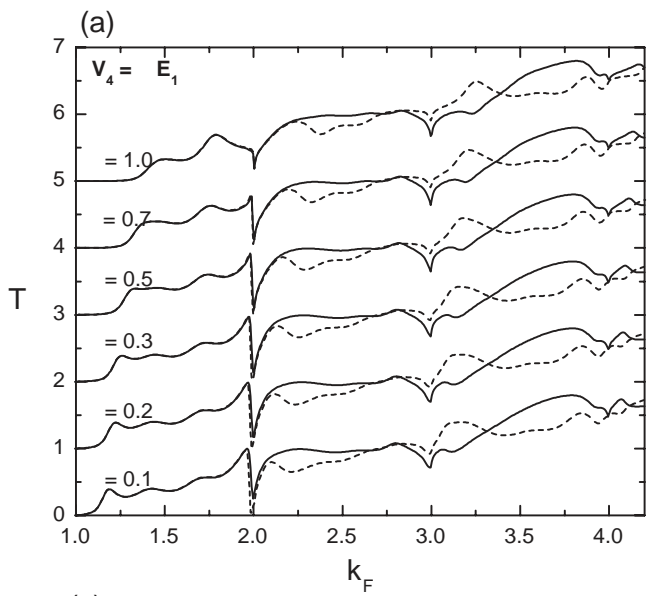
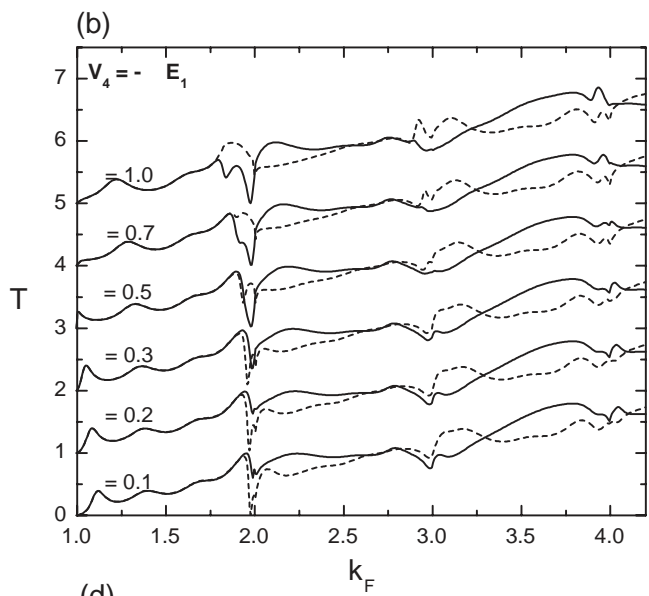
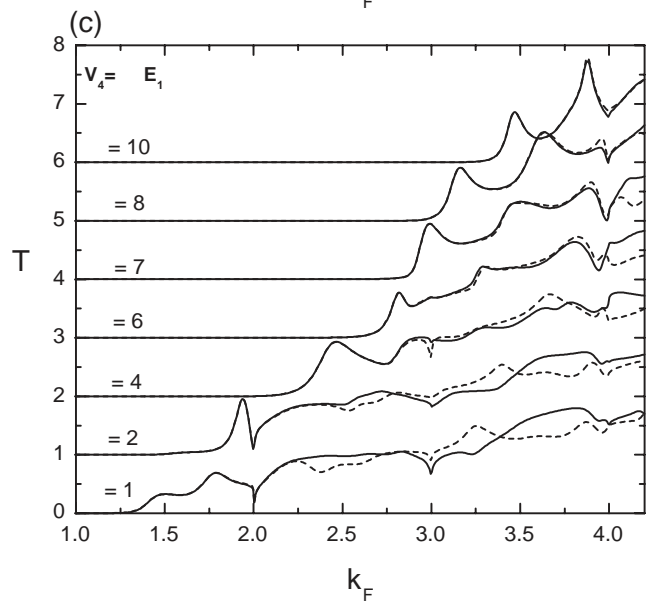
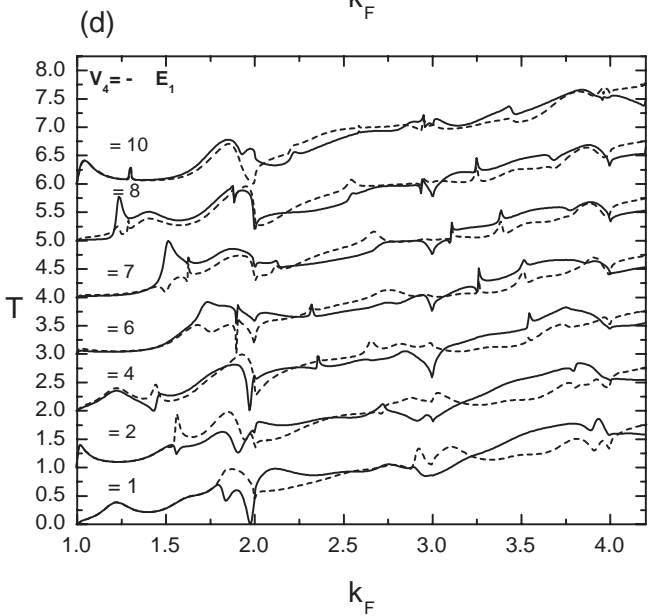
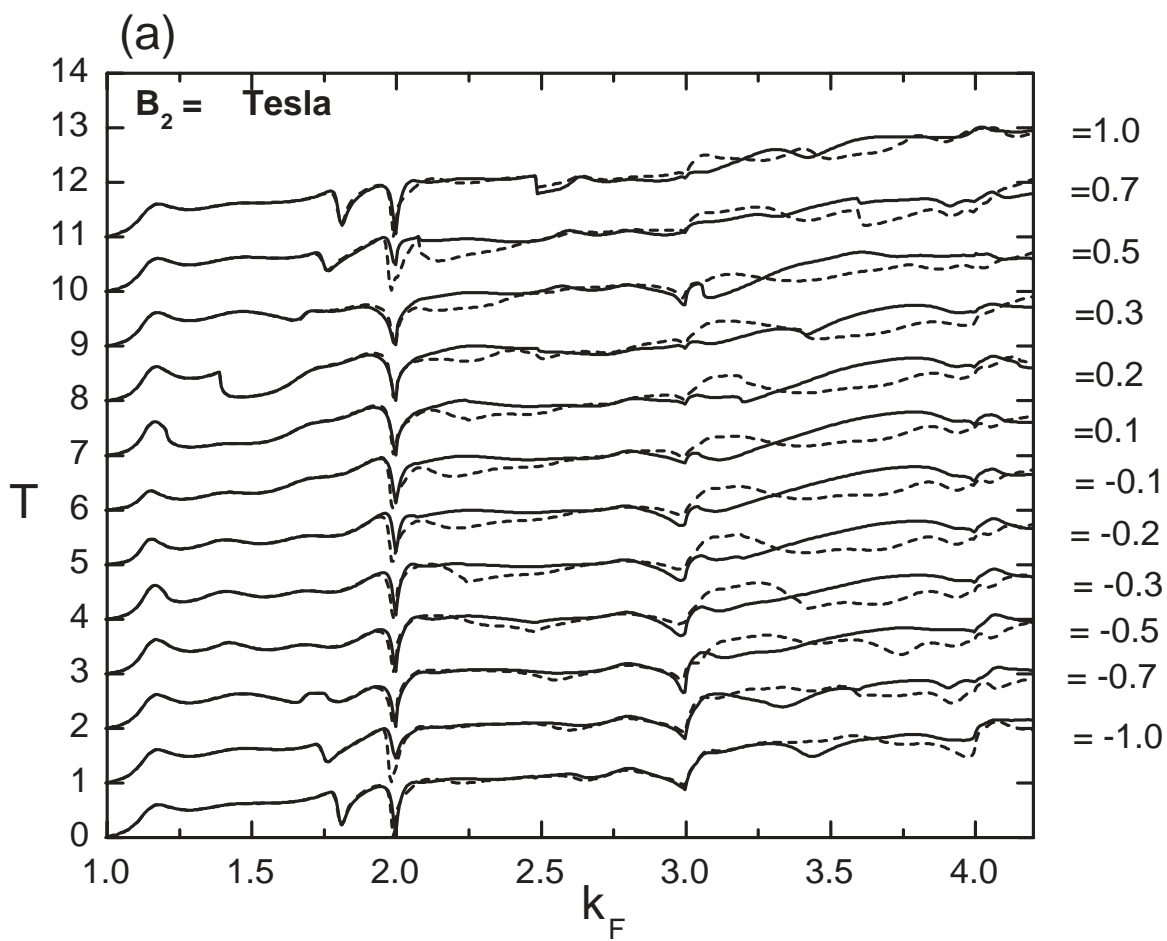
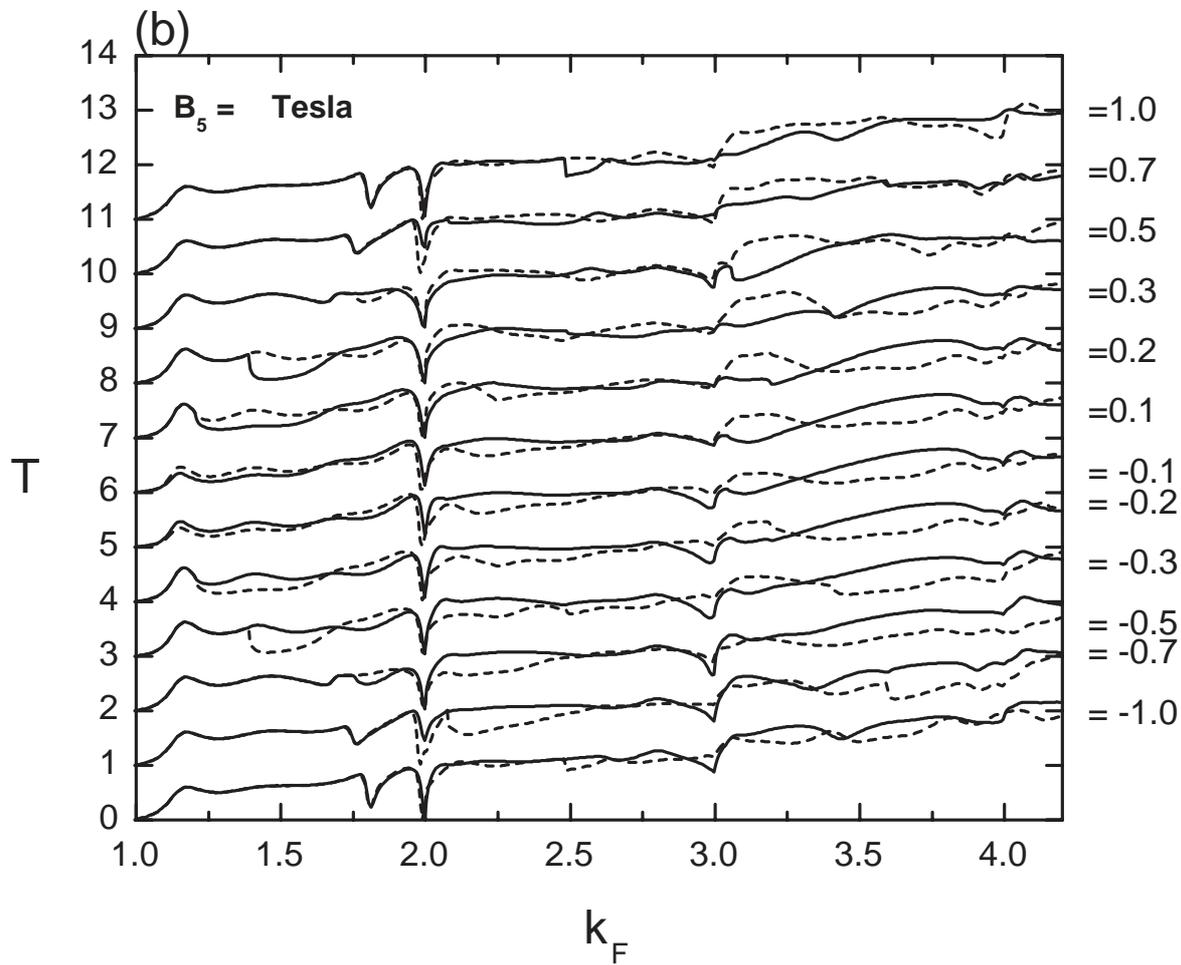
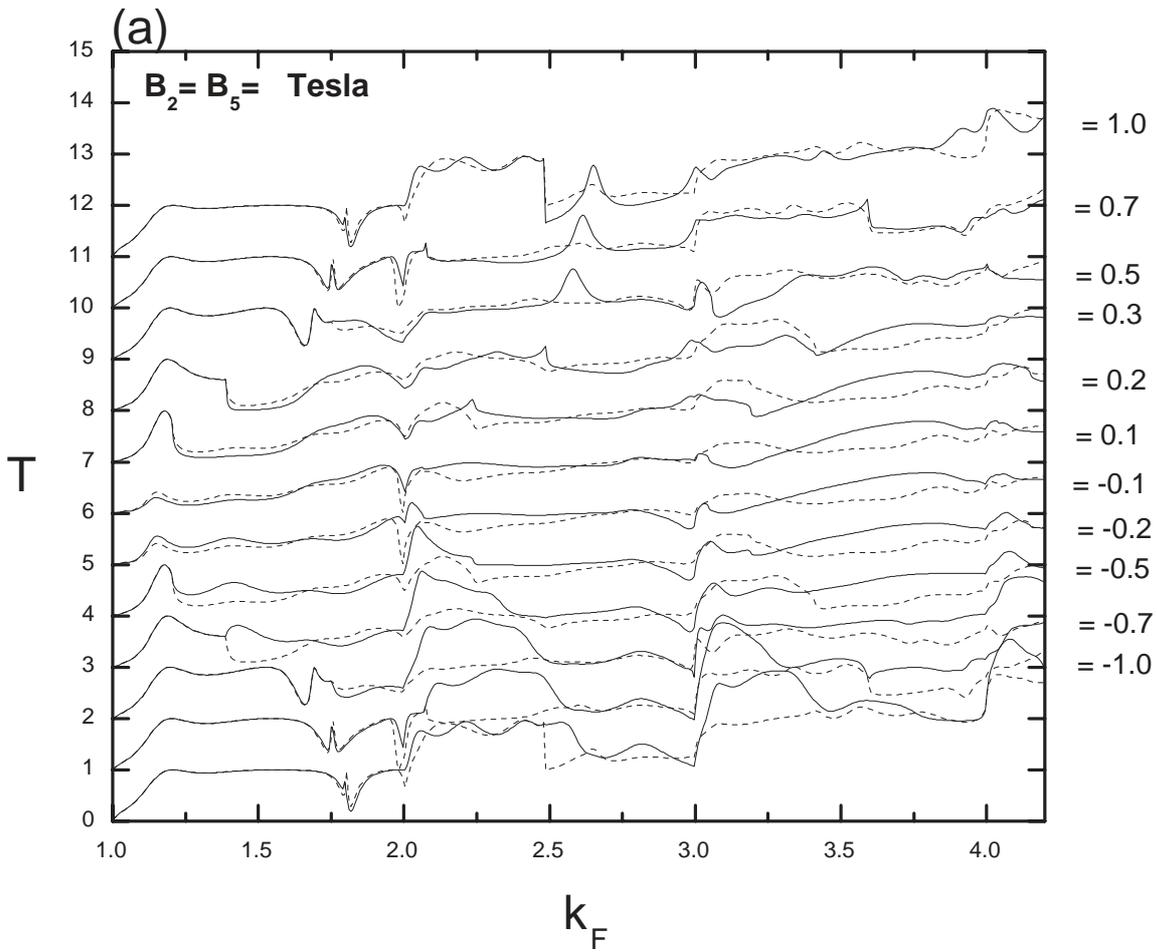
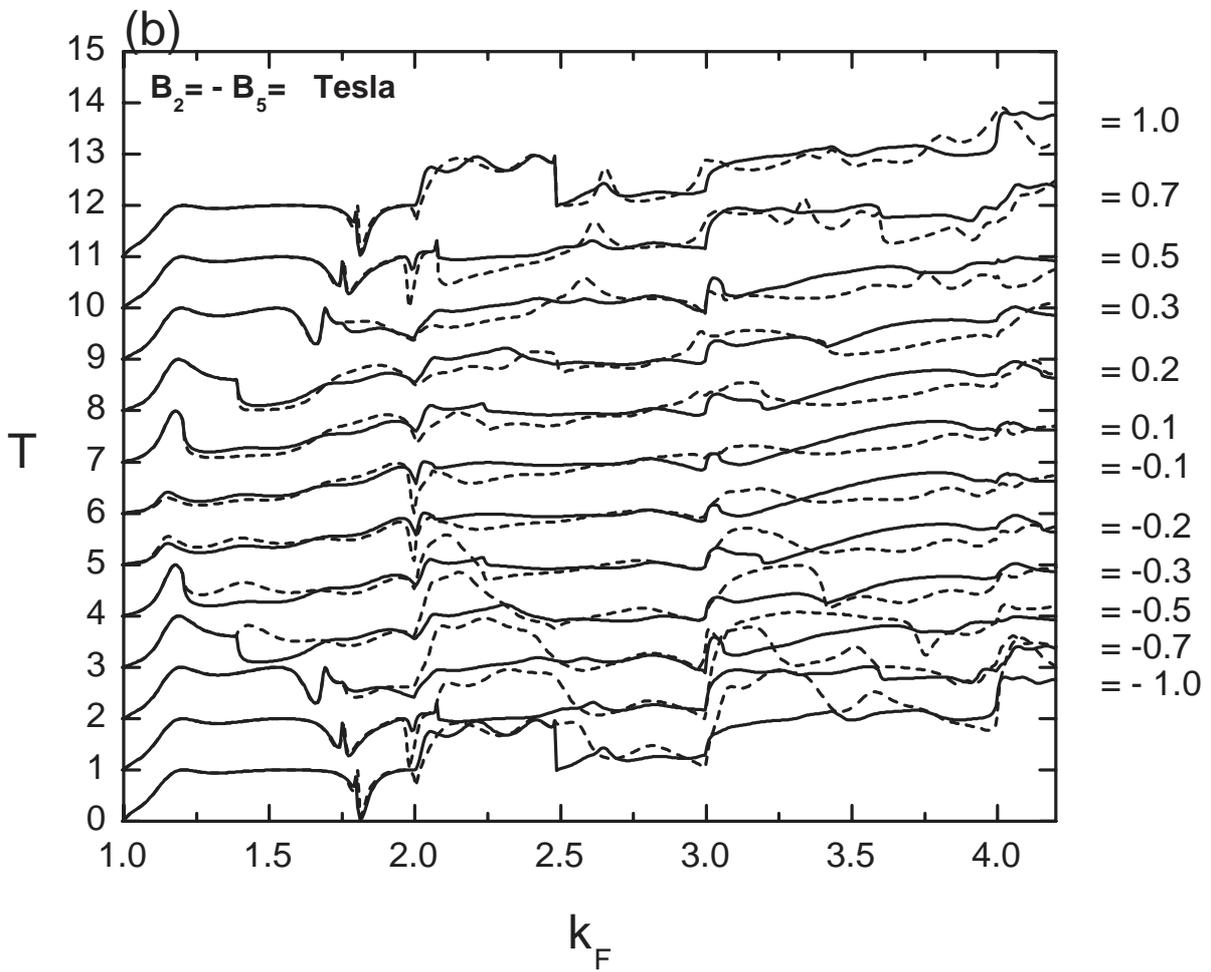
\end{document}